\documentclass[%
 reprint,
 amsmath,amssymb,
 aps,
]{revtex4-1}

\usepackage{graphicx}
\usepackage{dcolumn}
\usepackage{bm}

\usepackage[utf8]{inputenc}
\usepackage[T1]{fontenc}
\usepackage[french]{babel}
\usepackage{float}
\usepackage{gensymb} 
\usepackage{mathrsfs} 
\usepackage{braket}

\begin{document}


\title{Efficient quantum pseudorandomness with simple graph states}

\author{Rawad Mezher$^1$  $^2$}
 \email{rawad.mezher@etu.upmc.fr}
\author{Joe Ghalbouni$^1$}%
\email{joe.ghalbouni@ul.edu.lb}
\author{Joseph Dgheim$^1$}
\email{jdgheim@ul.edu.lb}
\author{Damian Markham$^2$}
 \email{damian.markham@lip6.fr}
\affiliation{(1) Laboratoire de Physique Appliquée, Faculty of Sciences 2, Lebanese University, 90656 Fanar, Lebanon\\
(2) LIP6-CNRS, Université Pierre et Marie Curie, Sorbonne Universit\'{e}s, 4 place Jussieu, 75252 Paris Cedex 05, France}

%
%

\date{\today}

\begin{abstract}
Measurement based (MB) quantum computation allows for universal quantum computing by measuring individual qubits prepared in entangled multipartite states, known as graph states. 
Unless corrected for, the randomness of the measurements leads to the generation of ensembles of random unitaries, where each random unitary is identified with a string of possible measurement results. 
We show that repeating an MB scheme an efficient number of times, on a simple graph state, with measurements at fixed angles and no feed-forward corrections, produces a random  unitary ensemble that is an $\varepsilon$-approximate t-design on n-qubits. 
Unlike previous constructions, the graph is regular and is also a universal resource for measurement based quantum computing, closely related to the brickwork state.
\end{abstract}

\maketitle


\section{\label{sec:level1}Introduction}

Randomness plays a prominent role in quantum computing, quantum information processing and physics in general. In particular, random unitaries chosen from the Haar measure \cite{4} on the unitary group U(N)  find  applications  in  randomized benchmarking \cite{6} , noise estimation \cite{5}, quantum metrology \cite{OAC+16}, as well as modeling thermalization \cite{7} and even black hole physics \cite{8}. 
Unfortunately, genuine Haar distributed unitaries are hard to create as the scaling required is exponential in the number of qubits \cite{9}. 
On  the other  hand   efficient  substitutes of  Haar distributed  unitaries   were  shown to  exist \cite{1}, \cite{10}, \cite{11}, \cite{12}, \cite{13}, \cite{3}, \cite{15}. These substitutes are known as unitary t-designs \cite{10} - ensembles over subsets of U(N) which mimic  exactly \cite{10}, \cite{11} , \cite{15} or approximately \cite{1}, \cite{12}, \cite{13},\cite{3},\cite{15} choosing from the Haar measure up to order t in the statistical moments. 

Following \cite{15} our approach is to harness the quantum randomness arising from applying a measurement based (MB) scheme  to produce approximate designs.
In MB computation \cite{14}, unitary determinsitic computation is achieved by making sequential, adaptive measurements on an entangled multipartite state, known as a graph state \cite{20}. Without these adaptive feedforward corrections, the inherent randomness of the measurements effectively samples from ensembles of unitaries.
In \cite{15} it was shown that starting with a fixed graph state, applying fixed angle measurements (with no need for feedforward corrections), effectively samples from an approximate t-design. 
Furthermore this process is efficient in the number of qubits, preperation and measurements, following from the efficiency of the construction of Brandao et al. \cite{1}. 
Indeed the construction of the graph state essentially mimics the random circuit construction of Brandao et al. \cite{1}.
However, in doing so, the graph itself is rather complicated, and moreover is not a simple regular lattice. 
A natural question is then, can simple, regular lattices (such as those useful for universal measurement based computation \cite{14}, \cite{19})   applied to generate t-designs?
As well as being more convenient from a practical point of view (in terms of generating the graph state), this connects the question of optimal generation of ensembles to standard measurement based quantum computation. Furthermore it requires a new proof that it is an approximate t-design (though the techniques also follow along the lines of \cite{1} it does not follow directly from their results).

In this  work we  show that it is possible. In particular we show that running fixed measurement MB scheme on a regular graph with poly-log (in n,  t and $\dfrac{1}{\varepsilon}$  ) number of qubits, with no feed-forward, results in an ensemble of random unitaries which forms a $\varepsilon$-approximate t-design ensemble. 
The graph we use is very similar to the brickwork graph known to be a universal resource for MB quantum computation \cite{19}. The proofs presented here rely principally on the G-local random circuit construction (GLRC) of Brandao et al. \cite{1}, the  detectability lemma (DL) of Aharonov et al. \cite{2} as well as a theorem \cite{33}, \cite {3}, \cite{1} on the equivalence between tensor product expanders (TPE’s) and approximate t-designs.\\

This paper is divided as follows: section II defines some preliminary notions, section III provides a brief statement of the results, section IV contains a detailed proof of the results, finally section V discusses briefly some potential applications.

\section{\label{sec:level2}Preliminaries}

\subsection{\label{subsec:level1} MBQC}

The model of computation used throughout this work is the measurement based quantum computation (MBQC) model. This model was first proposed by Raussendorf and Briegel \cite{14} as an alternative to the gate model of quantum computing \cite{16}.
Computation is carried out by first preparing a large entangled state (a graph state), followed by single qubit measurements. 
Crucially, in order to deterministically perform a desired unitary, the measurements must be done adaptively - to counter the inherent randomness arising from the measurements the measurement angles are corrected by feedforward process using previous measurement results.  However, instead of doing these corrections we will use this randomness as a resource to sample from an ensemble of unitaries with the desired structure - namely that they are a t-design.

A \emph{graph state} \cite{20} is a pure entangled quantum state of $n$ qubits in one to one correspondence with a graph $G=\{E,V\}$ of $n$ vertices  and edges. Each vertex $i \in V$ of the graph is associated with a qubit, and each edge ${i,j \in E}$ represents a preperation entanglement
\begin{eqnarray}
|G\rangle = \prod_{i,j \in E} CZ_{i,j} |+...+\rangle_V \nonumber.
\end{eqnarray}
For computations on quantum inputs, a set of vertices $I \subset  V$ are assigned as the input vertices, with initial input state $|\psi_{in}\rangle_I$, and the associated \emph{open graph state} is defined as
\begin{eqnarray} \label{eq Open Graph State}
|G(\psi)\rangle = \prod_{i,j \in E} CZ_{i,j} |\psi_{in}\rangle_I|+...+\rangle_{V\setminus I}.
\end{eqnarray}
We also identify the set of output qubits by the vertices $O\subset V$. Computation is then carried out by sequentially measuring all the non output qubits, in our case in a basis on the $X-Y$ plane. Each measurement is represented by an angle $\alpha$, corresponding to measureing in the basis \newline $\{|\pm\alpha\rangle := (|0\rangle \pm e^{i \alpha}|1\rangle)\}$.
In standard MBQC the correction strategy is given by the gflow \cite{21}- a partial order over the graph as well as a function, which give a time order for the measurements and the dependency respectively. In this work, following \cite{15},  we do not adapt the measurement angles so that for each measurement result a potentially different unitary is performed.

\begin{figure}[h]
\begin{center}

\graphicspath{}

\includegraphics[trim={0 10cm 0 0cm}, scale=0.18]{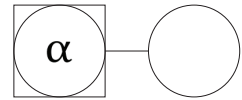}
\caption{MB scheme on a 2 qubit cluster state, the input (squared) qubit when measured at an angle $\alpha$ in the XY plane results in propagation of the input state to output along with application of a random unitary $U$ =$ H $$Z^{m}$ $Z$($\alpha$).}
\label{fig1}
\end{center}
\end{figure}

Following the convention of \cite{21}  (see also \cite{15}) we represent these resources as graphs where the input vertices have squares on them and the measured qubits have an angle inside representing the measurement angle, hence the quantum outputs $O$ are empty circles (note that in \cite{21} measured qubits are simply coloured black).
Figure \ref{fig1} illustrates this for a simple example. Following equation (\ref{eq Open Graph State}), for input state $|\psi_{in}\rangle = a|0\rangle+b|1\rangle$ Fig. \ref{fig1} corresponds to an initial open graph state 
\begin{eqnarray}
|G_\psi\rangle & =& a |0\rangle |+\rangle + b |1\rangle |-\rangle \nonumber \\
& = &\frac{1}{\sqrt{2}}\left(|+\alpha\rangle HZ(\alpha) |\psi_{in}\rangle +|-\alpha\rangle H Z Z(\alpha) |\psi_{in}\rangle \right)\nonumber,
\end{eqnarray}
where $H$ is the Hadamard gate, $Z$ is the Pauli Z gate, $Z(\alpha):=e^{-i \alpha Z/2}$ is a rotation by angle $\alpha$ around the Z axis. Denoting $m$ as the binary measurement outcome, associating $m=0$ to outcome $+\alpha$ and $m=1$ to $-\alpha$, it is clear that measuring the first qubit is equivalent to applying the random unitary
\begin{eqnarray}
U(m) = H Z^m Z(\alpha), 
\end{eqnarray}
with equal probabilty for $m=0$ and $m=1$.
As in \cite{15}, the same idea applied to larger graphs, with more inputs and outputs is the source of the random unitary ensembles we will study in this work.

\subsection{\label{subsec:level2}t-designs and Tensor Product Expanders}
Unitary t-designs \cite{10}, \cite{12}, \cite{13}  are subsets of the unitary group U(N) , sampled with some probability distribution, which mimic either exactly or approximately choosing from the Haar measure on U(N) up to order t in the statistical moments. Our concern in this work is with the latter, known as $\varepsilon$-approximate t-designs. 
More formally, let H =($C ^{ 2} )^{ \otimes n} $  be a Hilbert space on n qubits, define the density matrix $\rho$=$\ket{\phi}\bra{\phi}$ with $\ket{\phi}$  a unit vector in H (n-qubit state). Then let \{ $p_{i}$ , $U_{i}$ \} be a collection of unitaries in U(N)= U($2^{n}$) (the n-qubit unitary group), where we sample each $ U_{i}$ with probability $p_{i}$ . Let $\mu_{H}$ denote the Haar measure \cite{4}  on U($2^{n}$ ). Consider now t-copies of our n-qubit system, an ensemble $\{p_i, U_i\}$ is an  \emph{exact t-design} if it satisfies
\begin{equation}
\begin{aligned}
\label{eq1}
\sum_{i}p_{i} U_{i} ^{\otimes t} \rho^{t} U_{i} ^{\dagger \otimes t}=\int_{ U(2^{n})}U ^{\otimes t}\rho^{t}U ^{\dagger \otimes t}\mu_{H}(dU),
\end{aligned}
\end{equation}
for all $\rho^{t} \in \mathcal{B}(H^{\otimes t})$, where $\int_{ U(2^{n})}U ^{\otimes t}\rho^{t}U ^{\dagger \otimes t}\mu_{H}(dU)$ represents averaging over the Haar measure. 


$\varepsilon$-approximate t-designs are defined similarly as ensembles which satisfy
\begin{equation}
\begin{aligned}
\label{eq2}
(1-\varepsilon)\int_{ U(2^{n})}U ^{\otimes t}\rho^{t}U ^{\dagger \otimes t}\mu_{H}(dU) \leq \sum_{i}p_{i} U_{i} ^{\otimes t} \rho^{t} U_{i} ^{\dagger \otimes t} \\ \leq (1+\varepsilon) \int_{ U(2^{n})}U ^{\otimes t}\rho^{t}U ^{\dagger \otimes t}\mu_{H}(dU).
\end{aligned}
\end{equation}

Approximate t-designs can also be defined using various norms \cite{1}, \cite{13}, \cite{3}.
A useful concept equivalent to a t-design is a tensor product expander (TPE) \cite{22} . We say that a couple \{ $p_{i}$ , $U_{i}$ \}  is an ($\eta$,t)-TPE if the following equation holds \cite{3}, \cite{22} :

\begin{equation}
\begin{aligned}
\label{eq3}
g(t,\mu):=\mid\mid
\sum_{i }p_i U_i^{\otimes t,t} - \int_{U(2^{n}) } U^{\otimes t,t}\mu_{H}(dU) \mid\mid_\infty\leq\eta,
\end{aligned}
\end{equation}
where $U^{\otimes t,t} =U^{\otimes t} \otimes U^{\star \otimes t}$,  $\star$ denotes the complex conjugate and $\mu$ represents the probability measure on U(d) which results in choosing $U_i$ with probability $p_i$. 
In our main proof we will rely on the following theorem \cite{33}, \cite {3}, \cite{1} on the equivalence between TPE’s and $\varepsilon$-approximate t-designs:

\newtheorem{theorem}{Theorem}
\begin{theorem} \cite{33}, \cite {3}, \cite{1}\\
 Let \{ $p_{i}$ , $U_{i}$ \} be an ($\eta$,t)-TPE with $\eta$  < 1.Denote by $\mathcal{U}_{i}$ the set of all possible unitaries $U_{i}$.  Then iterating this TPE k times ( i.e. obtaining the product $U$=$\prod_{j=1,...,k}$$U_{\pi (j)}$  with the $U_{\pi (j)}$'s independently chosen from the ensemble  \{$p_{i}$ , $U_{i}$\} ,\\ $\pi$(j) $\in$ $\{1,…,|\mathcal{U}_{i}|\}$ with k $\ge$ $\dfrac{1}{log(\dfrac{1}{\eta})}log(\dfrac{d^{t}}{\varepsilon})$ results in an ensemble\{ {$p_U$  , $U$=$\prod_{j=1,...,k}$$U_{\pi (j)}$  }\} which is an $\varepsilon$-approximate t-design. Here $d$ is the dimension of the unitary group.
\end{theorem}

For convenience we define the t’th \emph{moment super operator} of  \{$p_{i}$ , $U_{i}$\}  (or simply moment super operator) as follows:
\begin{equation}
\begin{aligned}
\label{momop}
  M_t [\mu]:=\sum_{i} p_i U_i^{\otimes t,t}.
\end{aligned}
\end{equation}
Its role in the TPE condition (\ref{eq3}) means  $M_t [\mu]$  plays a major role in our proofs. Indeed, for any ensemble such that sampling it many times does eventually lead to the Haar measure, then $g(t,\mu)$ is equal to the second highest eigenvalue of $M_t[\mu]$ \cite{12,1}.
We will then follow the techniques of  \cite{1}, \cite{12}, \cite{13} in connecting the calculation of this to gaps of Hamiltonians, which will allow us to prove the connection to t-designs via theorem 1.

\subsection{\label{subsec:level3}Many body Physics and t-designs}

It has been known for some time \cite{1}, \cite{12}, \cite{13}  that the problem of estimating the scaling rate (number of iterations needed to reach a desired accuracy $\varepsilon$) of an $\varepsilon$- approximate t-design can be reduced to a problem of finding the spectral gap (the difference of energy between the ground and first excited state) of some many-body Hamiltonian. 
Here we give an overview of these techniques, in particular as used in \cite{1}. 

\begin{figure}[h]
\begin{center}

\graphicspath{}
\includegraphics[trim={0 12cm 0 16cm}, scale=0.08]{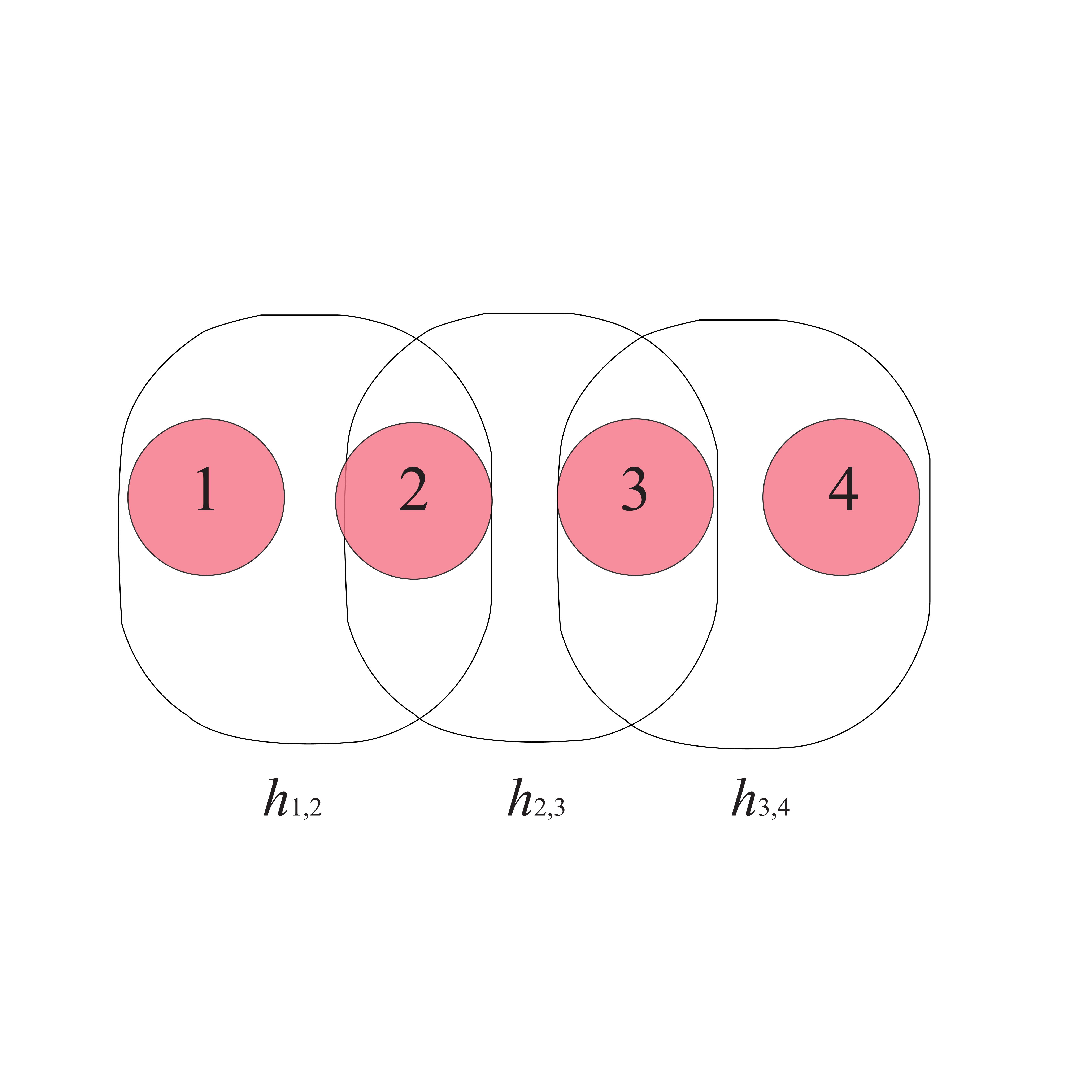}
\caption{Example of a 4 spins 1D system. The local Hamiltonians $h_{i,i+1}$  have a range of 2 (i.e. act on nearest neighbor spins). For example $h_{1,2}$ acts on spins 1 and 2. Translational invariance means that any $ h_{i,i+1}$ has the same form on all 2 qubit systems ($i$, $i$+1). In this case the total Hamiltonian of the system of 4 spins can be written as a sum of local Hamiltonians, i.e. H=$h_{1,2}$+$h_{2,3}$+$h_{3,4}$. }
\label{fig2}
\end{center}
\end{figure}

An extensive body of research (\cite{23}, \cite{24}, \cite {25}, \cite{26} and many others) has been devoted to the case of 1D spin chains, with local Hamiltonians (we assume a finite interaction range) with translational symmetry (see Fig.\ref{fig2}). 
We will focus exclusively on this case, and more precisely on a type of 1D Hamiltonian (the one we use in our proof) consisting of local terms acting on nearest neighbor spins $ i$ and $i$+1 with translational symmetry, which are frustration free (the entire Hamiltonian can be minimized by minimizing each of its local terms individually) as well as verifying the Nachtergaele criterion (\cite{23}, condition C.3). 
This family of  Hamiltonians was used by Brandao et al. \cite {1}  to study their local random circuit (LRC) construction, and later the so-called G-local random circuits (GLRC). 
We will briefly define these families of circuits and review these proofs.

The local random circuits (LRC) in \cite{1} generate random circuits on n-qubits as follows.
For each run of the LRC, a unitary $ U$ $\in$ U(4) is chosen from the Haar measure on U(4), then an index $i$  is chosen uniformly at random from the set \{1,…., n-1\} , finally $U$ is applied to qubits $i$ and $ i$+1. 
The LRC defines a couple \{$\mu_{LRC}$,$\mathcal{U}$\} where $\mu_{LRC}$ is the probability measure induced by one LRC run, and $\mathcal{U}$ is the set of  all the possible unitaries which can be generated by one LRC run. 
We arrive at the following moment super operator associated to one run of the LRC
\begin{equation}
\begin{aligned}
\label{eq4}
M_{t}[\mu_{LRC}]=\dfrac{1}{n-1}\sum_{i=1}^{n-1}\int_{U(4)}U_{i,i+1}^{\otimes t,t}\mu_{H}(dU)=\dfrac{1}{n-1}\sum_{i=1}^{n-1}P_{i,i+1},
\end{aligned}
\end{equation}
\\ where $ U_{i,i+1}$=$ 1^{\otimes i-1}\otimes U \otimes1^{\otimes n-i-1}$, $U$ $ \in$ U(4), $P_{i,i+1}:=\int_{U(4)}U_{i,i+1}^{\otimes t,t}\mu_{H}(dU)$ and $\mu_{H}$ is the Haar measure on U(4). 

 Since each of the $P_{i,i+1}$’s is Hermitian, then $M_t [\mu_{LRC} ]$  is itself Hermitian. 
Now consider the Hamiltonian 
\begin{equation}
H=\sum_{i}h_{i,i+1},
\end{equation}
where $ h_{i,i+1}$=$1- P_{i,i+1}$. Then
\begin{equation}
M_{t}[\mu_{LRC}] = I-\dfrac{H}{n-1}
\end{equation} 
The ground space of H has an eigenvalue of 0 and the gap between its ground and first excited spaces gives the second highest eigenvalue of $M_{t}[\mu_{LRC}]$.
This $H$ is a 1D spin chain Hamiltonian with nearest neighbor local terms which are translationally invariant. It is also frustration free by construction because the Hamiltonian can be minimized simply by minimizing all of its local terms (taking their ground state of energy 0) individually.
Brandao et al. also proved that this Hamiltonian verifies the Nachtergaele criterion, then also bounded the Nachtergaele Bound using path coupling techniques \cite{1}. 
In this way they show that the spectral gap $\Delta H$  of $H$ admits the following (polynomial in t) bound for n $\ge$ $\lfloor 2.5log_{2}(4t) \rfloor$ :
\begin{equation}
\begin{aligned}
\label{eq5}
\Delta H \ge (1700. \lfloor log_{2} (4t) \rfloor^{2}. t^{5}. t^{\dfrac{3.1}{log(2)}})^{-1},
\end{aligned}
\end{equation}
where $\lfloor x \rfloor$ denotes the floor function acting on variable $x$.

We now move to G-local random circuits, which are the finite set counter parts of LRC.
One run of the GLRC follows exactly as the LRC case, but instead of choosing a unitary $U$ from the Haar measure of U(4), we choose with uniform probability from a finite set G of SU(4) which is universal and contains inverses. One can show from the beautiful result of Bourgain and Gamburd \cite{26} that the Hamiltonian
$H_{GLRC}$=$\sum_{i} h^{'}_{i,i+1}$=$\sum_{i}(1-P^{'}_{i.i+1})$ with 
$P^{'}_{i.i+1}$ =$\dfrac{1}{|G|} \sum_{U \in G}(1^{\otimes i-1}\otimes U \otimes1^{\otimes n-i-1})^{\otimes t,t}$  admits the following bound for its spectral gap:
\begin{equation}
\begin{aligned}
\label{eq6}
\Delta H_{GLRC} \ge \alpha.\Delta H,
\end{aligned}
\end{equation}
with $\alpha$ a constant and $\Delta H$ the spectral gap of the LRC Hamiltonian.

Note that because the set G contains unitaries and their inverses and samples them uniformly, then $ \mu_{G }$($U$)=$\mu_{G}$ ($U^{\dagger}$),  for all $U$ ,$ U^{\dagger}$ $\in$ G. This means that $P^{'}_{i,i+1}$ (hence $ H_{GLRC}$ ) a Hermitian operator, and the above definition of a GLRC Hamiltonian makes sense.
We can rewrite equation (\ref{eq6}) as follows:
\begin{equation}
\begin{aligned}
\label{eq7}
\Delta H_{GLRC} \ge (C. \lfloor log_{2} (4t)\rfloor^{2}. t^{5}. t^{\dfrac{3.1}{log(2)}})^{-1}= P_{GLRC}
\end{aligned}
\end{equation}
 C being a constant depending on the gate set G. 

These spectral gaps directly give the second highest eigenvalue of the corresponding moment super operators, equal to the $g(t, \mu)$ in equation (\ref{eq3}) confirming the TPE conditions, which through theorem 1 then allow statements about their efficiency as t-designs \cite{1}.

\section{\label{sec:level4} Main Results}

\begin{figure}[h]
\begin{center}

\graphicspath{}
\includegraphics[trim={0 12cm 0 0cm}, scale=0.4]{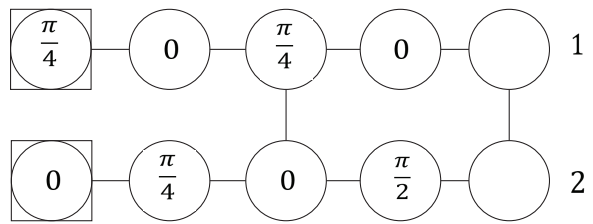}
\caption{The 2-row, 5-column brickwork state gadget giving rise to   $S_{I_{1 }}$ }
\label{fig3}
\end{center}
\end{figure}

\begin{figure}[h]
\begin{center}

\graphicspath{}
\includegraphics[trim={0 12cm 0 0cm}, scale=0.4]{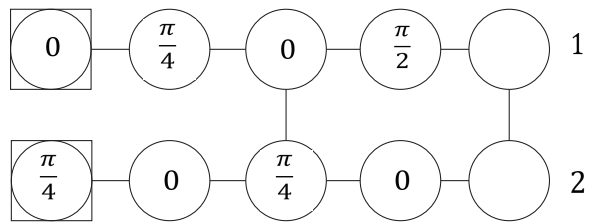}
\caption{The 2-row, 5-column brickwork state gadget giving rise to   $S_{I_{2 }}$ }
\label{fig4}
\end{center}
\end{figure}

In order to state the main results, we define some simple graph states for two input qubits. We will  hence forth refer to a graph state with an MB scheme applied to it as a gadget. These will act as the building blocks for our construction. 
Consider the 2, 5-column 2-row brickwork states with a fixed angle MB scheme in Fig.\ref{fig3} and Fig.\ref{fig4} which we call $S_{I_{1}}$  and $S_{I_{2}}$ .


$S_{I_{1} }$  and $ S_{I_{2} }$ give rise respectively to 2 MB ensembles \{$\dfrac{1}{2^{8}}$,$U_{1_{i}}$\} and \{$\dfrac{1}{2^{8}}$,$U_{2_{i}}$\}. 
The number of unitaries generated by the MB scheme on the 2-row, 5-column brickwork state is $2^{8}$.

It can be easily checked that each unitary of the ensembles $S_{I_{1} }$  and $ S_{I_{2} }$  contains (up to a global phase) an inverse in the ensemble. 
That is, denoting $\mathcal{U}_{S_{I_1}}$ as the set of unitaries generated by $S_{I_1}$ (and similarly for $S_{I_2}$),  for all $U_{1_{i}}$$\in \mathcal{U}_{S_{1_i}}$
 there exists $U_{1_{j}}$$\in \mathcal{U}_{S_{I_1}}$ such that $U_{1_{i}}=U^{\dagger}_{1_{j}}$. Simlarly for $S_{I_2}$.

Consider now a 13-column brickwork gadget: B= $S_{I_{1}} \circ S_{I_{2 }} \circ S_{I_{1} }$, where we mean by W $\circ$ V   a concatenation which identifies the output of graph W as an input of graph V.  We are now in a position to state our main results:\\ \\

\begin{theorem}
  The gadget B gives rise to an ensemble of unitaries which is \\
  (i) universal on SU(4) \\
  (ii) contains elements and their inverses, and\\
  (iii) is sampled with a uniform probability. \\
\end{theorem}

This theorem means that the set of unitaries generated by B (call it $\mathcal{U}_B$) satisfies the conditions necessary to form a GLRC. Though this is not exactly how our construction works, it will be important in proving our construction works in the proof of theorem 3.

\begin{figure}[h]
\begin{center}

\graphicspath{}
\includegraphics[trim={0 5cm 0 0cm},scale=0.4]{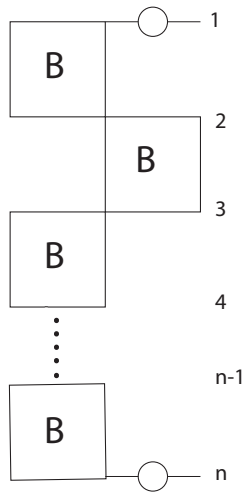}
\caption{The graph gadget $G_{n}$ pictured here for even n (the odd n case follows straightforwardly) }
\label{fig5}
\end{center}
\end{figure}

Now consider the gadget on n-qubits given in Fig.\ref{fig5} which we call $G_{n}$. 
The horizontal line with a circle in the middle means a direct link between output and input performed only on the 1st and last rows.  
The square with the letter B  is our 13-column brickwork gadget B, and the empty 3 sided square means that there is no vertical entanglement.    

The first and last rows  of $G_{n}$ are made up of 13 qubits, and all rows in between are made up of 25 qubits. This gives rise in total to a graph composed of  26$+$25(n$-$2) = 25n$-$24 qubits. We now state our second main result. \\ \\

\begin{theorem}
 The k(n,t,$\varepsilon$)-fold concatenation of $G_{n}$,  \newline $E_{n}$ = $G_{n} \circ G_{n} \circ$... , results in an ensemble of unitaries which forms an $\varepsilon$-approximate t-design on n-qubits ( n $\ge$ $\lfloor 2.5log_{2}(4t) \rfloor$ ), with: \\  k(n,t,$\varepsilon$)  $\ge$  $\dfrac{3}{log_{2}(1+\dfrac{P_{GLRC}}{2})}(nt+log_{2}(\dfrac{1}{\varepsilon}))$
\end{theorem}

\section{\label{sec:level5}Proofs}
\subsection{\label{subsec:level6}Proof of Theorem 2}
Before going on to universality, let us briefly explain why the set of unitaries generated by B,  $\mathcal{U}_B$ contains inverses ($(ii)$ in Theorem 2). 
Any element $U$ $\in$ $\mathcal{U}_B$ may be written as: $U$=$U_{1}.U_{2}.U_{1}^{'}$, where $U_{1}$  , $U_{1}^{'}$ $\in$ $\mathcal{U}_{S_{I_1}}$and $U_{2}$ $\in$ $\mathcal{U}_{S_{I_2}}$. 
Since $\mathcal{U}_{S_{I_1}}$and $\mathcal{U}_{S_{I_2}}$contain unitaries and their inverses, we can always find $ U^{\dagger}$=$U_{1}^{ ' \dagger}$.$U_{2}^{\dagger}$.$U_{1}^{\dagger}$  $\in$ $\mathcal{U}_B$. Furthermore, each unitary $U_{\{m\}} \in \mathcal{U}_B$ associated to a specific binary string $\{m\}$ is sampled with a uniform probability of $\dfrac{1}{|\mathcal{U}_B|}$, proving $(iii)$ in Theorem 2.


The remainder of this subsection is devoted to proving universality ($(i)$ in Theorem 2), and we will use the approach outlined in \cite{27} and \cite{34} for doing so. Following  \cite{27} and \cite{34}, one can show that the group generated by the set of unitaries \{$A$,$A^{\dagger}$,$C$,$C^{\dagger}$,$E$,$E^{\dagger}$,$F$,$F^{\dagger}$\} is dense (universal) on U(4) if the following conditions are satisfied:\\ \\
$C_{1}$ :   $H_{1}$:=$\dfrac{log(A)}{i}$ , $H_{2}$:=$\dfrac{log(C)}{i}$, $H_{3}$:=$\dfrac{log(E)}{i}$ and $H_{4}$:=$\dfrac{log(F)}{i}$  and their commutators form a set of 16 linearly independent Hamiltonians which span the Lie algebra \cite{28} of U(4).\\ \\
$C_{2}$  :  $H_{1}$, $ H_{2}$, $ H_{3}$ and $ H_{4}$ have eigenvalues that are irrationally related to $\pi$ .\\

We first consider $C_{1}$. We found 4 distinct unitaries $A$=$U_{\{m\}}$, $C$=$U_{\{m^{'}\}}$, $E$=$U_{\{m^{''}\}}$, and $F$=$U_{\{m^{'''}\}}$ .  Where $U_{\{m\}}$, $U_{\{m^{'}\}}$, $U_{\{m^{''}\}}$ and $U_{\{m^{'''}\}}$ $\in$ $\mathcal{U}_B$  are associated to the binary strings \{$m$\}=\{0,1,1,0,1,1,1,0,0,1,1,0,0,0,0,0,0,1,1,0,1,1,1,0\} , \{$m^{'}$\}=\{0,0,0,1,1,1,1,0,1,0,1,0,1,1,1,1,0,1,1,1,0,1,0,0\}, \{$m^{''}$\}=\{0,0,0,1,1,1,1,0,1,0,1,0,1,1,1,1,0,1,1,1,0,1,0,0\}, and \{$m^{'''}$\}=\{0,1,1,1,0,1,1,0,0,0,0,1,0,0,0,1,0,0,1,0,0,0,1,0\}. We adopt the convention that  the first 12 binary numbers appearing  in a given binary string represent the measurement results on qubits of the first row of B from left to right (input towards output), and the last 12 binaries represent the measurements performed on the qubits of the second row of B from left to right.

We then construct 16 Hamiltonians $H_{1}$ ,…, $H_{16}$ as follows:\\ \\
$H_{1}=\dfrac{log(A)}{i}$\\
$H_{2}=\dfrac{log(C)}{i}$\\
$H_{3}=\dfrac{log(E)}{i}$\\
$H_{4}=\dfrac{log(F)}{i}$\\
$H_{5}=i.[H_{1},H_{2}]$\\
$H_{6}=i.[H_{1},H_{3}]$\\
$H_{7}=i.[H_{1},H_{4}]$\\
$H_{8}=i.[H_{2},H_{3}]$\\
$H_{9}=i.[H_{2},H_{4}]$\\
$H_{10}=i.[H_{2},H_{5}]$\\
$H_{11}=i.[H_{2},H_{6}]$\\
$H_{12}=i.[H_{3},H_{4}]$\\
$H_{13}=i.[H_{3},H_{5}]$\\
$H_{14}=i.[H_{3},H_{6}]$\\
$H_{15}=i.[H_{4},H_{5}]$\\
$H_{16}=i.[H_{4},H_{6}]$\\

After that, we expand each of the 16 Hamiltonians in the basis:
 P=\{$P_{ij}$\} $i$,$j$=0,..,3, where  P is a basis of the Lie algebra of U(4) . 
 In other words, we write each: $H_{k}$=$a_{k}^{ij}$.($P_{ij}$) (Einstein summation convention adopted over i and j), where the  $a_{k}^{ij}$  ’ s are real numbers. 
 Since P is a basis of the Lie algebra of U(4) (over the field of real numbers ),  proving linear independence of the 16 Hamiltonians \{$H_{k}$\} $k$=1,..16  in the basis P means that the set \{$H_{k}$\} is itself a basis of the Lie algebra of U(4). 
 The linear independence of the 16 generators \{$H_k$\} is equivalent to the non-vanishing of the determinant of a 16 by 16 matrix M, where each of the 16 columns of M are made up of the 16 coefficients\{$a_{k}^{ij}$\} for a given $ k$. 
We found that the 16 Hamiltonians of our above constructed scheme give rise to a matrix M with non-vanishing determinant \footnote{this was done numerically however well within numerical precision}, thus this scheme forms a basis of the Lie algebra of U(4) and $C_{1}$ is verified for a subset of $\mathcal{U}_B$ (and hence for $\mathcal{U}_B$ itself ). 

Proving $C_{2}$ requires the use of a result in algebraic number theory called Lehmer’s theorem \cite{30}. Its context is described in the following Lemma:
\newtheorem{lem}{Lemma}
\begin{lem} \cite{30} 
  If n>2 and k and n are coprime integers, then 2cos($\dfrac{2k\pi}{n}$) is an algebraic integer.
\end{lem}
An algebraic number  is a complex number which is a solution of a polynomial equation with integer coefficients. The minimal polynomial of an algebraic number $z$ is the polynomial of lowest degree with integer coefficients for which $z$ is a solution. An algebraic integer is an algebraic number whose minimal polynomial is monic (that is, the coefficient in front of the highest degree variable is 1) \cite{31}. 

Lehmer’s theorem states that angles $\alpha$ = $\dfrac{2k\pi}{n}$ which are rationally related to $\pi$ must have $2cos(\alpha)$  an algebraic integer. So, if we can find instances of angles $\alpha$ in which $2cos(\alpha)$ is not an algebraic integer, then $\alpha$ has to be irrationally related to $\pi$  as a consequence of Lehmer’s theorem. 
Each of the eigenvalues $\lambda$ of  $A$,  $C$, $E$ or $F$ is a complex number with unit norm (because they are unitary matrices). Thus, $\lambda$=$e^{i\theta}$. We  calculated the expression $ 2cos(\theta)$
and constructed its minimal polynomial. 
We found that for each of the eigenvalues $\lambda$ of $A$ , $C$, $E$ and $F$, $ 2cos(\theta)$  does not verify a monic minimal polynomial (not an algebraic integer) 
and thus all the $\theta$’s are irrationally related to $\pi$ by Lehmer’s theorem. Further, because $A$, $C$, $E$ and $F$ are diagonal in the same basis as their Hamiltonians $H_{1}$, $H_{2}$, $H_{3}$ and $H_{4}$ \cite{27}, 
  the $\theta$’s we calculated are the eigenvalues of these Hamiltonians. Hence, the eigenvalues of the Hamiltonians are irrationally related to $\pi$ which proves $C_{2}$. 

Proving $C_{1}$ and $C_{2}$ means that the subset  \{$A$,$A^{\dagger}$,$C$,$C^{\dagger}$,$E$,$E^{\dagger}$,$F$,$F^{\dagger}$\} of $\mathcal{U}_B$ is universal on U(4), and thus so is $\mathcal{U}_B$. But $(i)$ further requires that the set be on SU(4). Fortunately, the moment super operator of a set sampled from U(4) can always be thought of as a sampling from SU(4). This can be seen by noting that for all $U$ $\in$ U(4) we have $det(U)$ $\neq$ 0, hence $U^{\otimes t,t }$=$|det(U)|^{\dfrac{t}{2}}$.$U^{ ' \otimes t,t}$=$U^{ ' \otimes t,t}$,  where $U^{'}$ $\in$ SU(4).

\subsection{\label{subsec:level7}Proof of Theorem 3}
Our approach for proving Theorem 3 can be summarized by 2 steps.  
In the first step, we prove that the ensemble generated by the gadget $G_{n}$ is an ($\eta$,t)-TPE with $\eta$=poly(t) <1 . We do so by using Aharonov et al.’s  detecability lemma \cite{2}. 
The second step uses Theorem 1 to establish the bound on $k(n,t,\varepsilon)$.

Consider a GLRC n-qubit Hamiltonian
\begin{eqnarray}
H_{GLRC} &=& \sum_{i}h^{'}_{i,i+1} \nonumber \\
&=&\sum_{i}(1-P^{'}_{i,i+1})
\end{eqnarray}
with G = $\mathcal{U}_B$, and  \\ $P^{'}_{i,i+1}$=$\dfrac{1}{|\mathcal{U}_B|} \sum_{U\in \mathcal{U}_B}(1^{\otimes i-1}\otimes U\otimes1^{\otimes n-i-1})^{\otimes t,t}$. Define $ P_{odd}$=$P^{'}_{1,2} .P^{'}_{3,4}$…
and  $P_{even}$=$P^{'}_{2,3} . P^{'}_{4,5}$….$ P_{odd}$ and $ P_{even}$ can be considered as projectors onto the "odd" and "even" ground spaces of  $H_{GLRC}$. Let $P_{0}$ be the projector onto the entire ground space of  $H_{GLRC}$. Further, because $H_{GLRC}$ is constructed from  universal sets on U(4), then its ground space projector is nothing but the  t'th  Haar moment super operator \cite{12}, 
In other words $P_{0}$=$\int_{U(2^{n})}$$U^{\otimes t,t }$$ \mu_{H}(dU)$  ;   $U$ $\in$ U($2^{n}$) and $\mu_{H}$ being the Haar measure on U($2^{n}$) .

The statement of the detecability lemma is the following :
\begin{lem} \cite{2} \\ $\mid\mid P_{even}.P_{odd} - P_{0} \mid\mid_{\infty}$  $\leq$  $(1+\dfrac{\Delta H_{GLRC}}{2} )^{-\dfrac{1}{3}}$
\end{lem}

To relate this to the ensemble generated by the gadget $ G_{n}$ we prove the following Lemma:
\begin{lem}
 $ M_{t} [\mu_{G_{n} } ]$=$P_{even}.P_{odd}$
\end{lem}
\textbf{Proof of Lemma 3}  :\\ \\ We first note that because all unitaries are drawn independently, we can think of the moment super operator as being composed of 2 layers (an odd layer (left part of the gadget of  Fig.\ref{fig5}) and an even layer (right part of the gadget of  Fig.\ref{fig5} ), this is similar to reasoning found in \cite{3}). Then:\\ 	

$M_{t} [\mu_{G_{n} } ]$=($\dfrac{1}{|\mathcal{U}_B|})^{\delta_{even}}\sum_{U_{23} \in \mathcal{U}_B,U_{45} \in \mathcal{U}_B,...} (U_{23}\otimes U_{45} \otimes …)^{\otimes t,t}$. ($\dfrac{1}{|\mathcal{U}_B|})^{\delta_{odd}}\sum_{U_{12} \in \mathcal{U}_B,U_{34} \in \mathcal{U}_B,...} (U_{12} \otimes  U_{34} \otimes …)^{\otimes t,t}$, \\  
where  $\delta_{odd}$=\{ $\dfrac{n}{2}$  if  n$mod$2=0 or $\dfrac{n-1}{2}$  if n$ mod$2 =1 \} \\  and  $\delta_{even}$=\{ $\dfrac{n}{2}-1$  if n$mod$2=0 or $\dfrac{n-1}{2}$  if n$mod$2=1 \}.\\

Note that since the $U_{ii+1}$ ' s are independently drawn from $\mathcal{U}_B$, one can rewrite this as:\\ 

$M_{t} [\mu_{G_{n} } ]$= ($\dfrac{1}{|\mathcal{U}_B|}  \sum_{U_{23} \in \mathcal{U}_B}(1 \otimes U_{23} \otimes 1^{\otimes n-3})^{\otimes t,t}$ .\\$\dfrac{1}{|\mathcal{U}_B|}(\sum_{U_{45} \in \mathcal{U}_B}(1^{\otimes 3}\otimes U_{45}\otimes 1^{\otimes n-5})^{\otimes t,t}$...) .\\($\dfrac{1}{|\mathcal{U}_B|} (\sum_{U_{12} \in \mathcal{U}_B}(U_{12} \otimes 1^{\otimes n-2})^{\otimes t,t}$.\\$\dfrac{1}{|\mathcal{U}_B|}\sum_{U_{34} \in \mathcal{U}_B}(1^{\otimes 2} \otimes U_{34} \otimes 1^{\otimes n-4})^{\otimes t,t}$...)\\=$(P^{'}_{2,3}.P^{'}_{4,5}...)$.$(P^{'}_{1,2}.P^{'}_{3,4}...)$\\=$P_{even}$. $P_{odd}$. $\Box$ \\ \\  

Then, as a direct consequence of the detectibility lemma we obtain:
\begin{equation}
\begin{aligned}
\label{dl}
g(t,\mu_{G_{n} }) = \mid\mid M_{t} [\mu_{G_{n} } ] - P_{0} \mid\mid_{\infty}\leq(1+\dfrac{\Delta H_{GLRC}}{2} )^{-\dfrac{1}{3}}
\end{aligned}
\end{equation}
All that remains now is to bound the RHS of Equation (\ref{dl}). Using Equation (\ref{eq7}) one directly obtains:
\begin{equation}
\begin{aligned}
\label{eq8}
(1+\dfrac{\Delta H_{GLRC}}{2})^{-\dfrac{1}{3}}\leq(1+\dfrac{P_{GLRC}}{2})^{-\dfrac{1}{3}} 
\end{aligned}
\end{equation}
Equation (\ref{dl}) along with Equation (\ref{eq8}) directly leads to the following Corollary:
\newtheorem{corollary}{Corollary}
\begin{corollary}
The ensemble $\{\dfrac{1}{|\mathcal{U_{G}}|} , \mathcal{U_{G}}\}$ generated by the gadget $G_{n}$ is an ($\eta$,t)-TPE with :\\
 $\eta$=$(1+\dfrac{P_{GLRC}}{2})^{-\dfrac{1}{3}}$= poly(t) < 1.
\end{corollary}
Plugging Corollary 1 into Theorem 1 with: \\ \{ $p_{i}$, $U_{i}$ \}=$\{\dfrac{1}{|\mathcal{U_{G}}|} , \mathcal{U_{G}}\}$ ,  $d$=$2^{n}$ and $\eta$=$(1+\dfrac{P_{GLRC}}{2})^{-\dfrac{1}{3}}$ , then multiplying and dividing the bound of  $k$ in Theorem 1 by $log(2)$  allows one to obtain Theorem 3.

\section{\label{sec:level8}Conclusions and discussion}
We have found a simple n-qubit graph gadget which implements an $\varepsilon$-approximate t-design under repeated concatenations with fixed measurement and no feedforward.
The number of concatenations $k(n,t,\varepsilon)$=$\Omega(nt+ log(\dfrac{1}{\varepsilon}))$ required is linear in both the qubit number and order t of the design. Also, because the number of qubits in the  graph gadget scales linearly with  n, we thus   only require $\Omega(n^{2}t + nlog(\dfrac{1}{\varepsilon}))$ qubits in total to implement the gadget $E_{n}$=$G_{n} \circ G_{n} \circ...$. Furthermore, the choice of the 2-qubit gadget B is not at all unique. In fact, $G_{n}$ could be made even more practical provided simpler (less number of qubits, less needed entanglements,...) 2-qubit  gadgets possessing the properties of B can be found. 


Our construction is very similar to the brickwork state, which is a universal resource for MBQC \cite{19} - it is basically the brickwork state but with regular holes.
In MBQC these holes would simply teleport the inputs through, so that the proofs of universality of \cite{19} easily extend to our graph - that is, concatenations of the graph used in  $G_n$ is also a universal resource for MBQC.
In addition to being pleasing from a practical point of view, this opens the door to applications of techniques for delegation of ensemble generation, as done for computation \cite{19,barz2013experimental}, and indeed the possibility to hide whether one is sampling unitaries or performing some deterministic computation.\\ \\ \\

\section{\label{sec:level8}Acknowledgements}

We thank Y. Nakata for fruitful discussions and  P. Turner for comments. R. Mezher Acknowledges funding from the Lebanese PhD grant $CNRS-L$/$UL$.
DM acknowledges support from ANR grant COMB.


\bibliography{zzz}

\begin{thebibliography}{33}%
\makeatletter
\providecommand \@ifxundefined [1]{%
 \@ifx{#1\undefined}
}%
\providecommand \@ifnum [1]{%
 \ifnum #1\expandafter \@firstoftwo
 \else \expandafter \@secondoftwo
 \fi
}%
\providecommand \@ifx [1]{%
 \ifx #1\expandafter \@firstoftwo
 \else \expandafter \@secondoftwo
 \fi
}%
\providecommand \natexlab [1]{#1}%
\providecommand \enquote  [1]{``#1''}%
\providecommand \bibnamefont  [1]{#1}%
\providecommand \bibfnamefont [1]{#1}%
\providecommand \citenamefont [1]{#1}%
\providecommand \href@noop [0]{\@secondoftwo}%
\providecommand \href [0]{\begingroup \@sanitize@url \@href}%
\providecommand \@href[1]{\@@startlink{#1}\@@href}%
\providecommand \@@href[1]{\endgroup#1\@@endlink}%
\providecommand \@sanitize@url [0]{\catcode `\\12\catcode `\$12\catcode
  `\&12\catcode `\#12\catcode `\^12\catcode `\_12\catcode `\%12\relax}%
\providecommand \@@startlink[1]{}%
\providecommand \@@endlink[0]{}%
\providecommand \url  [0]{\begingroup\@sanitize@url \@url }%
\providecommand \@url [1]{\endgroup\@href {#1}{\urlprefix }}%
\providecommand \urlprefix  [0]{URL }%
\providecommand \Eprint [0]{\href }%
\providecommand \doibase [0]{http://dx.doi.org/}%
\providecommand \selectlanguage [0]{\@gobble}%
\providecommand \bibinfo  [0]{\@secondoftwo}%
\providecommand \bibfield  [0]{\@secondoftwo}%
\providecommand \translation [1]{[#1]}%
\providecommand \BibitemOpen [0]{}%
\providecommand \bibitemStop [0]{}%
\providecommand \bibitemNoStop [0]{.\EOS\space}%
\providecommand \EOS [0]{\spacefactor3000\relax}%
\providecommand \BibitemShut  [1]{\csname bibitem#1\endcsname}%
\let\auto@bib@innerbib\@empty
\bibitem [{\citenamefont {De~Chiffre}(2011)}]{4}%
  \BibitemOpen
  \bibfield  {author} {\bibinfo {author} {\bibfnamefont {M.~D.}\ \bibnamefont
  {De~Chiffre}},\ }\emph {\bibinfo {title} {The Haar measure}},\ \href@noop {}
  {Ph.D. thesis},\ \bibinfo  {school} {Department of Mathematical Sciences,
  University of Copenhagen} (\bibinfo {year} {2011})\BibitemShut {NoStop}%
\bibitem [{\citenamefont {Epstein}\ \emph {et~al.}(2014)\citenamefont
  {Epstein}, \citenamefont {Cross}, \citenamefont {Magesan},\ and\
  \citenamefont {Gambetta}}]{6}%
  \BibitemOpen
  \bibfield  {author} {\bibinfo {author} {\bibfnamefont {J.}~\bibnamefont
  {Epstein}}, \bibinfo {author} {\bibfnamefont {A.~W.}\ \bibnamefont {Cross}},
  \bibinfo {author} {\bibfnamefont {E.}~\bibnamefont {Magesan}}, \ and\
  \bibinfo {author} {\bibfnamefont {J.~M.}\ \bibnamefont {Gambetta}},\
  }\href@noop {} {\bibfield  {journal} {\bibinfo  {journal} {Physical Review
  A}\ }\textbf {\bibinfo {volume} {89}} (\bibinfo {year} {2014})}\BibitemShut
  {NoStop}%
\bibitem [{\citenamefont {Emerson}\ \emph {et~al.}(2003)\citenamefont
  {Emerson}, \citenamefont {Weinstein}, \citenamefont {Saraceno}, \citenamefont
  {Lloyd},\ and\ \citenamefont {Cory}}]{5}%
  \BibitemOpen
  \bibfield  {author} {\bibinfo {author} {\bibfnamefont {J.}~\bibnamefont
  {Emerson}}, \bibinfo {author} {\bibfnamefont {Y.~S.}\ \bibnamefont
  {Weinstein}}, \bibinfo {author} {\bibfnamefont {M.}~\bibnamefont {Saraceno}},
  \bibinfo {author} {\bibfnamefont {S.}~\bibnamefont {Lloyd}}, \ and\ \bibinfo
  {author} {\bibfnamefont {D.~G.}\ \bibnamefont {Cory}},\ }\href {DOI:
  10.1126/science.1090790} {\bibfield  {journal} {\bibinfo  {journal}
  {Science}\ }\textbf {\bibinfo {volume} {302}},\ \bibinfo {pages} {2098}
  (\bibinfo {year} {2003})}\BibitemShut {NoStop}%
\bibitem [{\citenamefont {Oszmaniec}\ \emph {et~al.}(2016)\citenamefont
  {Oszmaniec}, \citenamefont {Augusiak}, \citenamefont {Gogolin}, \citenamefont
  {Ko{\l}ody{\'n}ski}, \citenamefont {Acin},\ and\ \citenamefont
  {Lewenstein}}]{OAC+16}%
  \BibitemOpen
  \bibfield  {author} {\bibinfo {author} {\bibfnamefont {M.}~\bibnamefont
  {Oszmaniec}}, \bibinfo {author} {\bibfnamefont {R.}~\bibnamefont {Augusiak}},
  \bibinfo {author} {\bibfnamefont {C.}~\bibnamefont {Gogolin}}, \bibinfo
  {author} {\bibfnamefont {J.}~\bibnamefont {Ko{\l}ody{\'n}ski}}, \bibinfo
  {author} {\bibfnamefont {A.}~\bibnamefont {Acin}}, \ and\ \bibinfo {author}
  {\bibfnamefont {M.}~\bibnamefont {Lewenstein}},\ }\href@noop {} {\bibfield
  {journal} {\bibinfo  {journal} {Physical Review X}\ }\textbf {\bibinfo
  {volume} {6}},\ \bibinfo {pages} {041044} (\bibinfo {year}
  {2016})}\BibitemShut {NoStop}%
\bibitem [{\citenamefont {Müller}\ \emph {et~al.}(2015)\citenamefont
  {Müller}, \citenamefont {Adlam}, \citenamefont {Masanes},\ and\
  \citenamefont {Wiebe}}]{7}%
  \BibitemOpen
  \bibfield  {author} {\bibinfo {author} {\bibfnamefont {M.}~\bibnamefont
  {Müller}}, \bibinfo {author} {\bibfnamefont {E.}~\bibnamefont {Adlam}},
  \bibinfo {author} {\bibfnamefont {L.}~\bibnamefont {Masanes}}, \ and\
  \bibinfo {author} {\bibfnamefont {N.}~\bibnamefont {Wiebe}},\ }\href@noop {}
  {\bibfield  {journal} {\bibinfo  {journal} {Communications in Mathematical
  Physics}\ }\textbf {\bibinfo {volume} {340}},\ \bibinfo {pages} {499}
  (\bibinfo {year} {2015})}\BibitemShut {NoStop}%
\bibitem [{\citenamefont {Hayden}\ and\ \citenamefont {Preskill}(2007)}]{8}%
  \BibitemOpen
  \bibfield  {author} {\bibinfo {author} {\bibfnamefont {P.}~\bibnamefont
  {Hayden}}\ and\ \bibinfo {author} {\bibfnamefont {J.}~\bibnamefont
  {Preskill}},\ }\href@noop {} {\bibfield  {journal} {\bibinfo  {journal}
  {Journal of High Energy Physics}\ }\textbf {\bibinfo {volume} {120}}
  (\bibinfo {year} {2007})}\BibitemShut {NoStop}%
\bibitem [{\citenamefont {Knill}(1995)}]{9}%
  \BibitemOpen
  \bibfield  {author} {\bibinfo {author} {\bibfnamefont {E.}~\bibnamefont
  {Knill}},\ }\href@noop {} {\bibfield  {journal} {\bibinfo  {journal}
  {arXiv:quant-ph/9508006}\ } (\bibinfo {year} {1995})}\BibitemShut {NoStop}%
\bibitem [{\citenamefont {Brandao}\ \emph {et~al.}(2012)\citenamefont
  {Brandao}, \citenamefont {Harrow},\ and\ \citenamefont {Horodecki}}]{1}%
  \BibitemOpen
  \bibfield  {author} {\bibinfo {author} {\bibfnamefont {F.}~\bibnamefont
  {Brandao}}, \bibinfo {author} {\bibfnamefont {A.~W.}\ \bibnamefont {Harrow}},
  \ and\ \bibinfo {author} {\bibfnamefont {M.}~\bibnamefont {Horodecki}},\
  }\href@noop {} {\bibfield  {journal} {\bibinfo  {journal} {arXiv preprint
  arXiv:1208.0692}\ } (\bibinfo {year} {2012})}\BibitemShut {NoStop}%
\bibitem [{\citenamefont {Dankert}\ \emph {et~al.}(2009)\citenamefont
  {Dankert}, \citenamefont {Cleve}, \citenamefont {Emerson},\ and\
  \citenamefont {Livine}}]{10}%
  \BibitemOpen
  \bibfield  {author} {\bibinfo {author} {\bibfnamefont {C.}~\bibnamefont
  {Dankert}}, \bibinfo {author} {\bibfnamefont {R.}~\bibnamefont {Cleve}},
  \bibinfo {author} {\bibfnamefont {J.}~\bibnamefont {Emerson}}, \ and\
  \bibinfo {author} {\bibfnamefont {E.}~\bibnamefont {Livine}},\ }\href@noop {}
  {\bibfield  {journal} {\bibinfo  {journal} {Physical Review A}\ }\textbf
  {\bibinfo {volume} {80}} (\bibinfo {year} {2009})}\BibitemShut {NoStop}%
\bibitem [{\citenamefont {Zhu}(2015)}]{11}%
  \BibitemOpen
  \bibfield  {author} {\bibinfo {author} {\bibfnamefont {H.}~\bibnamefont
  {Zhu}},\ }\href@noop {} {\bibfield  {journal} {\bibinfo  {journal} {arXiv
  preprint arXiv:1510.02619}\ } (\bibinfo {year} {2015})}\BibitemShut {NoStop}%
\bibitem [{\citenamefont {Brown}\ and\ \citenamefont {Viola}(2010)}]{12}%
  \BibitemOpen
  \bibfield  {author} {\bibinfo {author} {\bibfnamefont {W.~G.}\ \bibnamefont
  {Brown}}\ and\ \bibinfo {author} {\bibfnamefont {L.}~\bibnamefont {Viola}},\
  }\href@noop {} {\bibfield  {journal} {\bibinfo  {journal} {Physical review
  letters}\ }\textbf {\bibinfo {volume} {104}} (\bibinfo {year}
  {2010})}\BibitemShut {NoStop}%
\bibitem [{\citenamefont {Harrow}\ and\ \citenamefont
  {Low}(2009{\natexlab{a}})}]{13}%
  \BibitemOpen
  \bibfield  {author} {\bibinfo {author} {\bibfnamefont {A.}~\bibnamefont
  {Harrow}}\ and\ \bibinfo {author} {\bibfnamefont {R.~A.}\ \bibnamefont
  {Low}},\ }\href@noop {} {\bibfield  {journal} {\bibinfo  {journal}
  {Communications in Mathematical Physics}\ }\textbf {\bibinfo {volume} {291}}
  (\bibinfo {year} {2009}{\natexlab{a}})}\BibitemShut {NoStop}%
\bibitem [{\citenamefont {Nakata}\ \emph {et~al.}(2016)\citenamefont {Nakata},
  \citenamefont {Hirche}, \citenamefont {Koashi},\ and\ \citenamefont
  {Winter}}]{3}%
  \BibitemOpen
  \bibfield  {author} {\bibinfo {author} {\bibfnamefont {Y.}~\bibnamefont
  {Nakata}}, \bibinfo {author} {\bibfnamefont {C.}~\bibnamefont {Hirche}},
  \bibinfo {author} {\bibfnamefont {M.}~\bibnamefont {Koashi}}, \ and\ \bibinfo
  {author} {\bibfnamefont {A.}~\bibnamefont {Winter}},\ }\href@noop {}
  {\bibfield  {journal} {\bibinfo  {journal} {arXiv preprint arXiv:1609.07021}\
  } (\bibinfo {year} {2016})}\BibitemShut {NoStop}%
\bibitem [{\citenamefont {Turner}\ and\ \citenamefont {Markham}(2016)}]{15}%
  \BibitemOpen
  \bibfield  {author} {\bibinfo {author} {\bibfnamefont {P.}~\bibnamefont
  {Turner}}\ and\ \bibinfo {author} {\bibfnamefont {D.}~\bibnamefont
  {Markham}},\ }\href@noop {} {\bibfield  {journal} {\bibinfo  {journal}
  {Physical review letters}\ }\textbf {\bibinfo {volume} {116}} (\bibinfo
  {year} {2016})}\BibitemShut {NoStop}%
\bibitem [{\citenamefont {Raussendorf}\ and\ \citenamefont
  {Briegel}(2001)}]{14}%
  \BibitemOpen
  \bibfield  {author} {\bibinfo {author} {\bibfnamefont {R.}~\bibnamefont
  {Raussendorf}}\ and\ \bibinfo {author} {\bibfnamefont {H.~J.}\ \bibnamefont
  {Briegel}},\ }\href@noop {} {\bibfield  {journal} {\bibinfo  {journal}
  {Physical Review Letters}\ }\textbf {\bibinfo {volume} {86}} (\bibinfo {year}
  {2001})}\BibitemShut {NoStop}%
\bibitem [{\citenamefont {Hein}\ \emph {et~al.}(2006)\citenamefont {Hein},
  \citenamefont {Dür}, \citenamefont {Eisert}, \citenamefont {Raussendorf},
  \citenamefont {Nest},\ and\ \citenamefont {Briegel}}]{20}%
  \BibitemOpen
  \bibfield  {author} {\bibinfo {author} {\bibfnamefont {M.}~\bibnamefont
  {Hein}}, \bibinfo {author} {\bibfnamefont {W.}~\bibnamefont {Dür}}, \bibinfo
  {author} {\bibfnamefont {J.}~\bibnamefont {Eisert}}, \bibinfo {author}
  {\bibfnamefont {R.}~\bibnamefont {Raussendorf}}, \bibinfo {author}
  {\bibfnamefont {M.}~\bibnamefont {Nest}}, \ and\ \bibinfo {author}
  {\bibfnamefont {H.~J.}\ \bibnamefont {Briegel}},\ }\href@noop {} {\bibfield
  {journal} {\bibinfo  {journal} {arXiv preprint quant-ph/0602096}\ } (\bibinfo
  {year} {2006})}\BibitemShut {NoStop}%
\bibitem [{\citenamefont {Broadbent}\ \emph {et~al.}(2009)\citenamefont
  {Broadbent}, \citenamefont {Fitzsimons},\ and\ \citenamefont {Kashefi}}]{19}%
  \BibitemOpen
  \bibfield  {author} {\bibinfo {author} {\bibfnamefont {A.}~\bibnamefont
  {Broadbent}}, \bibinfo {author} {\bibfnamefont {J.}~\bibnamefont
  {Fitzsimons}}, \ and\ \bibinfo {author} {\bibfnamefont {E.}~\bibnamefont
  {Kashefi}},\ }\href@noop {} {\bibfield  {journal} {\bibinfo  {journal} {In
  Foundations of Computer Science, 2009. FOCS'09. 50th Annual IEEE Symposium
  on}\ } (\bibinfo {year} {2009})}\BibitemShut {NoStop}%
\bibitem [{\citenamefont {Aharonov}\ \emph {et~al.}(2011)\citenamefont
  {Aharonov}, \citenamefont {Arad}, \citenamefont {Vazirani},\ and\
  \citenamefont {Landau}}]{2}%
  \BibitemOpen
  \bibfield  {author} {\bibinfo {author} {\bibfnamefont {D.}~\bibnamefont
  {Aharonov}}, \bibinfo {author} {\bibfnamefont {I.}~\bibnamefont {Arad}},
  \bibinfo {author} {\bibfnamefont {U.}~\bibnamefont {Vazirani}}, \ and\
  \bibinfo {author} {\bibfnamefont {Z.}~\bibnamefont {Landau}},\ }\href@noop {}
  {\bibfield  {journal} {\bibinfo  {journal} {New journal of physics}\ }\textbf
  {\bibinfo {volume} {13}} (\bibinfo {year} {2011})}\BibitemShut {NoStop}%
\bibitem [{\citenamefont {Harrow}\ and\ \citenamefont
  {Low}(2009{\natexlab{b}})}]{33}%
  \BibitemOpen
  \bibfield  {author} {\bibinfo {author} {\bibfnamefont {A.~W.}\ \bibnamefont
  {Harrow}}\ and\ \bibinfo {author} {\bibfnamefont {R.~A.}\ \bibnamefont
  {Low}},\ }\href@noop {} {\bibfield  {journal} {\bibinfo  {journal} {In
  Approximation, Randomization, and Combinatorial Optimization 12th
  International Workshop}\ } (\bibinfo {year}
  {2009}{\natexlab{b}})}\BibitemShut {NoStop}%
\bibitem [{\citenamefont {Nielsen}\ and\ \citenamefont {Chuang}(2002)}]{16}%
  \BibitemOpen
  \bibfield  {author} {\bibinfo {author} {\bibfnamefont {M.~A.}\ \bibnamefont
  {Nielsen}}\ and\ \bibinfo {author} {\bibfnamefont {I.}~\bibnamefont
  {Chuang}},\ }\enquote {\bibinfo {title} {Quantum computation and quantum
  information},}\ \ (\bibinfo  {publisher} {Cambridge University Press},\
  \bibinfo {year} {2002})\BibitemShut {NoStop}%
\bibitem [{\citenamefont {Browne}\ \emph {et~al.}(2007)\citenamefont {Browne},
  \citenamefont {Kashefi}, \citenamefont {Mhalla},\ and\ \citenamefont
  {Perdrix}}]{21}%
  \BibitemOpen
  \bibfield  {author} {\bibinfo {author} {\bibfnamefont {D.~E.}\ \bibnamefont
  {Browne}}, \bibinfo {author} {\bibfnamefont {E.}~\bibnamefont {Kashefi}},
  \bibinfo {author} {\bibfnamefont {M.}~\bibnamefont {Mhalla}}, \ and\ \bibinfo
  {author} {\bibfnamefont {S.}~\bibnamefont {Perdrix}},\ }\href@noop {}
  {\bibfield  {journal} {\bibinfo  {journal} {New Journal of Physics}\ }\textbf
  {\bibinfo {volume} {9}} (\bibinfo {year} {2007})}\BibitemShut {NoStop}%
\bibitem [{\citenamefont {Hastings}\ and\ \citenamefont {Harrow}(2008)}]{22}%
  \BibitemOpen
  \bibfield  {author} {\bibinfo {author} {\bibfnamefont {M.~B.}\ \bibnamefont
  {Hastings}}\ and\ \bibinfo {author} {\bibfnamefont {A.~W.}\ \bibnamefont
  {Harrow}},\ }\href@noop {} {\bibfield  {journal} {\bibinfo  {journal} {arXiv
  preprint arXiv:0804.0011}\ } (\bibinfo {year} {2008})}\BibitemShut {NoStop}%
\bibitem [{\citenamefont {Nachtergaele}(1996)}]{23}%
  \BibitemOpen
  \bibfield  {author} {\bibinfo {author} {\bibfnamefont {B.}~\bibnamefont
  {Nachtergaele}},\ }\href@noop {} {\bibfield  {journal} {\bibinfo  {journal}
  {Communications in mathematical physics}\ }\textbf {\bibinfo {volume} {175}}
  (\bibinfo {year} {1996})}\BibitemShut {NoStop}%
\bibitem [{\citenamefont {Hastings}\ and\ \citenamefont {Koma}(2006)}]{24}%
  \BibitemOpen
  \bibfield  {author} {\bibinfo {author} {\bibfnamefont {M.~B.}\ \bibnamefont
  {Hastings}}\ and\ \bibinfo {author} {\bibfnamefont {T.}~\bibnamefont
  {Koma}},\ }\href@noop {} {\bibfield  {journal} {\bibinfo  {journal}
  {Communications in mathematical physics}\ }\textbf {\bibinfo {volume} {265}}
  (\bibinfo {year} {2006})}\BibitemShut {NoStop}%
\bibitem [{\citenamefont {Affleck}(1989)}]{25}%
  \BibitemOpen
  \bibfield  {author} {\bibinfo {author} {\bibfnamefont {I.}~\bibnamefont
  {Affleck}},\ }\href@noop {} {\bibfield  {journal} {\bibinfo  {journal}
  {Journal of Physics: Condensed Matter}\ }\textbf {\bibinfo {volume} {1}}
  (\bibinfo {year} {1989})}\BibitemShut {NoStop}%
\bibitem [{\citenamefont {Bourgain}\ and\ \citenamefont {Gamburd.}(2011)}]{26}%
  \BibitemOpen
  \bibfield  {author} {\bibinfo {author} {\bibfnamefont {J.}~\bibnamefont
  {Bourgain}}\ and\ \bibinfo {author} {\bibnamefont {Gamburd.}},\ }\href@noop
  {} {\bibfield  {journal} {\bibinfo  {journal} {preprint arXiv:1108.6264.}\ }
  (\bibinfo {year} {2011})}\BibitemShut {NoStop}%
\bibitem [{\citenamefont {Deutsch}\ \emph {et~al.}(1995)\citenamefont
  {Deutsch}, \citenamefont {Barenco},\ and\ \citenamefont {Ekert}}]{27}%
  \BibitemOpen
  \bibfield  {author} {\bibinfo {author} {\bibfnamefont {D.}~\bibnamefont
  {Deutsch}}, \bibinfo {author} {\bibfnamefont {A.}~\bibnamefont {Barenco}}, \
  and\ \bibinfo {author} {\bibfnamefont {A.}~\bibnamefont {Ekert}},\
  }\href@noop {} {\bibfield  {journal} {\bibinfo  {journal} {In Proceedings of
  the Royal Society of London A: Mathematical, Physical and Engineering
  Sciences}\ }\textbf {\bibinfo {volume} {449}} (\bibinfo {year}
  {1995})}\BibitemShut {NoStop}%
\bibitem [{\citenamefont {Lloyd}(1995)}]{34}%
  \BibitemOpen
  \bibfield  {author} {\bibinfo {author} {\bibfnamefont {S.}~\bibnamefont
  {Lloyd}},\ }\href@noop {} {\bibfield  {journal} {\bibinfo  {journal}
  {Physical Review Letters}\ }\textbf {\bibinfo {volume} {75}} (\bibinfo {year}
  {1995})}\BibitemShut {NoStop}%
\bibitem [{\citenamefont {Jacobson}(1979)}]{28}%
  \BibitemOpen
  \bibfield  {author} {\bibinfo {author} {\bibfnamefont {N.}~\bibnamefont
  {Jacobson}},\ }\enquote {\bibinfo {title} {Lie algebras.}}\ \ (\bibinfo
  {publisher} {Interscience Publishers},\ \bibinfo {year} {1979})\BibitemShut
  {NoStop}%
\bibitem [{Note1()}]{Note1}%
  \BibitemOpen
  \bibinfo {note} {This was done numerically however well within numerical
  precision}\BibitemShut {NoStop}%
\bibitem [{\citenamefont {Niven}(2005)}]{30}%
  \BibitemOpen
  \bibfield  {author} {\bibinfo {author} {\bibfnamefont {I.}~\bibnamefont
  {Niven}},\ }\enquote {\bibinfo {title} {Irrational numbers},}\ \ (\bibinfo
  {publisher} {Cambridge University Press},\ \bibinfo {year}
  {2005})\BibitemShut {NoStop}%
\bibitem [{\citenamefont {Lang}(2013)}]{31}%
  \BibitemOpen
  \bibfield  {author} {\bibinfo {author} {\bibfnamefont {S.}~\bibnamefont
  {Lang}},\ }\enquote {\bibinfo {title} {Algebraic number theory},}\ \
  (\bibinfo  {publisher} {Springer},\ \bibinfo {year} {2013})\BibitemShut
  {NoStop}%
\bibitem [{\citenamefont {Barz}\ \emph {et~al.}(2013)\citenamefont {Barz},
  \citenamefont {Fitzsimons}, \citenamefont {Kashefi},\ and\ \citenamefont
  {Walther}}]{barz2013experimental}%
  \BibitemOpen
  \bibfield  {author} {\bibinfo {author} {\bibfnamefont {S.}~\bibnamefont
  {Barz}}, \bibinfo {author} {\bibfnamefont {J.~F.}\ \bibnamefont
  {Fitzsimons}}, \bibinfo {author} {\bibfnamefont {E.}~\bibnamefont {Kashefi}},
  \ and\ \bibinfo {author} {\bibfnamefont {P.}~\bibnamefont {Walther}},\
  }\href@noop {} {\bibfield  {journal} {\bibinfo  {journal} {arXiv preprint
  arXiv:1309.0005}\ } (\bibinfo {year} {2013})}\BibitemShut {NoStop}%
\end{thebibliography}%

\end{document}